\documentclass{article}
\usepackage{spconf}
\usepackage{KalmanNet}
 \usepackage{cite}
\usepackage[bookmarks,colorlinks]{hyperref}

\usepackage{enumitem}

\title{Unsupervised Learned Kalman Filtering}
\name{Guy Revach, Nir Shlezinger, Timur Locher, Xiaoyong Ni, Ruud J. G. van Sloun, and Yonina C. Eldar
\thanks{
		G. Revach. T. Locher, and X. Ni are with the Institute for Signal and Information Processing (ISI), D-ITET, ETH Zürich, Switzerland (e-mail: \{grevach; tlocher\}@ethz.ch,  xiaoni@student.ethz.ch). 
		N. Shlezinger is with the School of ECE, Ben-Gurion University of the Negev, Beer Sheva, Israel (e-mail: nirshl@bgu.ac.il). 
		R. J. G. van Sloun is with the EE Dpt., Eindhoven University of Technology, and with Phillips Research, Eindhoven,  The Netherlands (e-mail: r.j.g.v.sloun@tue.nl).
		Y. C. Eldar is with the Faculty of Math and CS, Weizmann Institute of Science, Rehovot, Israel (e-mail: yonina@weizmann.ac.il).
		The authors thank Prof. Hans-Andrea Loeliger for his helpful comments and discussion.}}
\address{\vspace{-25mm}}
\begin{document}
\maketitle
%
%
\begin{abstract}
In this paper we adapt \acl{kn}, which is a recently proposed \ac{dnn}-aided system whose architecture follows the operation of the \acl{mb} \ac{kf}, to learn its mapping in an unsupervised manner, i.e., without requiring ground-truth states. The unsupervised adaptation is achieved by exploiting the hybrid \acl{mb}/\acl{dd} architecture of \acl{kn}, which internally predicts the next observation as the \ac{kf} does. These internal features are then used to compute the loss rather than the state estimate at the output of the system. With the capability of unsupervised learning, one can use \acl{kn} not only to track the hidden state, but also to adapt to variations in the \ac{ss} model. We numerically demonstrate that when the noise statistics are unknown, unsupervised \acl{kn} achieves a {similar} performance to \acl{kn} with supervised learning. We also show that we can adapt a pre-trained \acl{kn} to changing \ac{ss} models without providing additional data thanks to the unsupervised capabilities. 
\end{abstract}
%
%
\acresetall 
%
\vspace{-0.3cm}
\section{Introduction}\label{sec:intro}
\vspace{-0.1cm}
Real-time tracking of hidden state sequences from noisy observations plays a major role in many signal processing systems. Classic approaches are based on the \ac{kf} \cite{kalman1960new} and its variants \cite[Ch. 10]{durbin2012time}. These \ac{mb} techniques rely on accurate knowledge of an underlying statistical \ac{ss} model capturing the system dynamics, which may not be available in some applications, and tend to notably degrade in the presence of model mismatch. To cope with missing model parameters, data is commonly used for parameter estimation, followed by plugging in the missing parameters into the \ac{mb} \ac{kf} and its variants \cite{abbeel2005discriminative, xu2021ekfnet}.

The unprecedented success of \acl{dl} has spurred a multitude of \acp{dnn} based approaches for \ac{ss} model related tasks, that are optimized in an \acl{e2e} manner. This allows to achieve improved accuracy compared with \ac{mb} algorithms when applied in complex, poorly understood, and partially known dynamics, by learning to carry out the task directly from data. Notable approaches include \ac{dnn} feature extractors \cite{ zhou2020kfnet},  variational inference techniques \cite{krishnan2015deep, karl2016deep, fraccaro2017disentangled, naesseth2018variational, archer2015black, krishnan2017structured}, and the usage of \acp{rnn} \cite{haarnoja2016backprop,zheng2017state, coskun2017long, becker2019recurrent}. When the \ac{ss} model is partially known, one can benefit from the available knowledge by using the hybrid \ac{mb}/\ac{dd} \acl{kn} architecture proposed in \cite{KalmanNetTSPa} for the \acl{rt} filtering task, as a learned \ac{kf} via \ac{mb} deep learning \cite{shlezinger2020model}.

A key challenge in applying an \acl{e2e} \ac{dnn}-based filters, stems from their need to be trained using labeled data i.e., a large volume of pairs of noisy measurements and their corresponding ground-truth hidden state sequences from the underlying \ac{ss} model. Obtaining such ground-truth sequences may be costly, particularly in setups where the underlying dynamics, i.e., the \ac{ss} model, change over time. Previous works on  \ac{dnn}-based for \ac{ss} model related tasks in the unsupervised setup, focused mostly on the imputation task of filling in missing observations\cite{archer2015black, karl2016deep, fraccaro2017disentangled, naesseth2018variational}. This task notably differs from real-time state estimations, also known as filtering \cite[Ch. 4]{durbin2012time}.

In this work we extend \acl{kn} \cite{KalmanNetTSPa}, to learn its mapping in an unsupervised fashion by building upon its interpretable hybrid architecture, which learns to implement the \ac{kf} while preserving its structure. Specifically,
we define a loss measure that uses the noisy observations and their predictions taken from an internal feature of \acl{kn}.
We also propose a semi-supervised training method, which first trains \acl{kn} offline, and then adapts in an unsupervised online manner to dynamics that differ from the offline trained model without providing ground-truth data.  This mechanism results in \acl{kn} tracking not only the latent state, but also changes in the underlying \ac{ss} model. Our numerical evaluations demonstrate that the unsupervised \acl{kn} that does not have access to the noise statistics, approaches the \ac{kf} with full domain knowledge. Furthermore, its semi-supervised implementation allows to improve upon on its supervised counterpart due to the newly added  ability to track variations in the \ac{ss} model without requiring additional data. 

{The rest of this paper is organized as follows:} 
Section~\ref{sec:SysModel} formulates the \ac{ss} model and the problem. Section~\ref{sec:KalmanNet} presents unsupervised \acl{kn}, which is  evaluated in Section~\ref{sec:NumEval}. %
%
%
\section{System Model and Preliminaries}\label{sec:SysModel}
We review the \ac{ss} model and briefly recall the supervised \acl{kn}. For simplicity, we focus on linear \ac{ss} models, though the derivations can also be used for non-linear models in the same manner as the extended \ac{kf} is applied \cite[Ch. 10]{durbin2012time}, as we demonstrate in Section~\ref{sec:NumEval}.
\subsection{Problem Formulation}\label{ssec:ssmdl} 
%
We consider state estimation in \acl{dt}, linear, Gaussian \ac{ss} models.  Letting  $\gvec{x}_{t}$  denote the $m\times 1$ hidden state vector at time instance $t\in\gint$, which evolves in time via
%
\begin{align}\label{eqn:stateEvolution}
\gvec{x}_{t}&= 
\gvec{F}\cdot{\gvec{x}_{t-1}}+\gvec{w}_{t},& 
\gvec{w}_t\sim
\mathcal{N}\brackets{\gvec{0},\gvec{Q}},&
\quad
\gvec{x}_{t}\in\greal^m.
\end{align}
Here, $\gvec{F}$ is the $m\times m$ state evolution matrix, while $\gvec{w}_t$ is \ac{awgn} with covariance $\gvec{Q}$. The corresponding observation    $\gvec{y}_{t}$, is related to   $\gvec{x}_{t}$ via 
\begin{align}
\label{eqn:stateObservation}
\gvec{y}_{t}&=
\gvec{H}\cdot{\gvec{x}_{t}}+\gvec{v}_{t},& 
\gvec{v}_t\sim
\mathcal{N}\brackets{\gvec{0},\gvec{R}},&
\quad
\gvec{y}_{t}\in\greal^n,
\end{align}
 where $\gvec{H}$ is the $n\times m$ measurement matrix, and $\gvec{v}_t$ is an \ac{awgn} with covariance $\gvec{R}$.  We focus on the filtering problem, where one needs to track the  hidden state ${\gvec{x}}_t$ from a known initial state $\gvec{x}_0$. At each time instance $t$, the goal is to provide an instantaneous  estimate $\hat{\gvec{x}}_{t}$, based on the observations seen so far $\set{\gvec{y}_\tau}_{\tau=1}^t$. 
We consider scenarios where one has partial domain knowledge, such that the  statistics of the noises $\gvec{w}_t$ and $\gvec{v}_t$ are not known, {while} the matrices $\gvec{F}$ and $\gvec{H}$ are {known}. To fill the information gap we assume that we have access to an \emph{unlabeled} \acl{ds} containing a sequence of observations from which one has to learn to recover the hidden state.  
%
%
\subsection{Supervised \acl{kn}} \label{ssec:supervised}
{\acl{kn}} is a hybrid \ac{mb}/\ac{dd} implementation of the \ac{kf}. The latter utilizes full knowledge of the \ac{ss} model to estimate $\gvec{x}_t$, based on the current observed $\gvec{y}_t$ and the previous estimate $\hat{\gvec{x}}_{t-1}$. This is achieved by first predicting the next state and observation based solely on the previous estimate via  
\begin{subequations}\label{eqn:predict1}
\begin{align}\label{eqn:evol}
\hat{\gvec{x}}_{t\given{t-1}} &= 
\gvec{F}\cdot{\hat{\gvec{x}}_{t-1}},\\\label{eqn:obs}
\hat{\gvec{y}}_{t\given{t-1}} &=
\gvec{H}\cdot{\hat{\gvec{x}}_{t\given{t-1}}},
\end{align}
\end{subequations}
while computing the second-order moments of these estimates as 
$\mySigma_{t\given{t-1}} =
\gvec{F}\cdot\mySigma_{t-1}\cdot\gvec{F}^\top+\gvec{Q}$,  and $\gvec{S}_{t\given{t-1}} =
\gvec{H}\cdot\mySigma_{t\given{t-1}}\cdot\gvec{H}^\top+\gvec{R}$.
%
%
Next, the \ac{kf} computes the \ac{kg} $\Kgain_t$ as $\Kgain_{t}={\mySigma}_{t\given{t-1}}\cdot{\gvec{H}}^\top\cdot{\gvec{S}}^{-1}_{t\given{t-1}}$, which is used to update the estimation covariance ${\mySigma}_{t}=
{\mySigma}_{t\given{t-1}}-\Kgain_{t}\cdot{\mathbf{S}}_{t\given{t-1}}\cdot\Kgain^{\top}_{t}$, and provide the state estimate $\hat{\gvec{x}}_t$ via
\begin{equation}\label{eq:update1}
\hat{\gvec{x}}_{t}=
\hat{\gvec{x}}_{t\given{t-1}}+\Kgain_{t}\cdot\Delta\gvec{y}_t,
\end{equation}
where $\Delta\gvec{y}_t$ is the innovation process computed as
\begin{equation}\label{eqn:innov}
\Delta\gvec{y}_t=\gvec{y}_t-\hat{\gvec{y}}_{t\given{t-1}}.
\end{equation}
%
%

%
\begin{figure} 
\includegraphics[width=1\columnwidth]{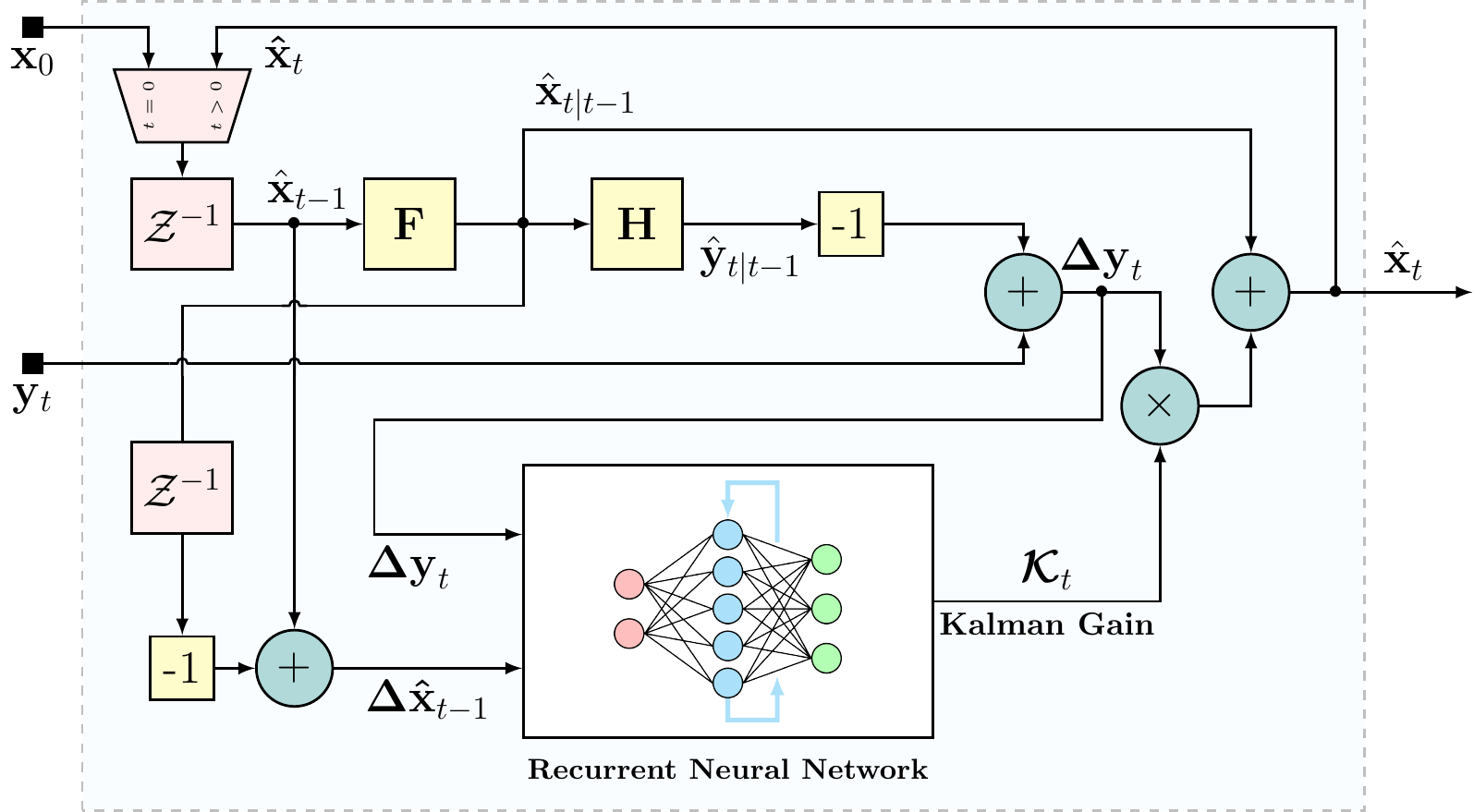}
\caption{\acl{kn} block diagram.}
\label{fig:KNet_Block}
\end{figure} 

\acl{kn} learns to implement the \ac{kf} from labeled data in partially known \ac{ss} models. This is achieved by noting that the available  knowledge allows to compute the predictions in \eqref{eqn:predict1}, while the missing domain knowledge is needed to compute the \ac{kg}. Thus, \acl{kn} augments the flow of the \ac{kf} with an \ac{rnn}, which estimates the \ac{kg} and implicitly tracks the second-order moments computed by the \ac{kf}, while the state estimate is obtained via \eqref{eq:update1}  (see \cite{KalmanNetTSPa} for a detailed description). The \acl{kn} architecture, depicted in Fig.~\ref{fig:KNet_Block}, is trained to estimate the state in a supervised manner based on labeled data. The data set comprises $N$ pairs of hidden state trajectories and its corresponding observations of the form $\mathcal{D} = \set{\brackets{\gvec{Y}_i, \gvec{X}_{i}}}_{i=1}^N$, where
\begin{align}
\gvec{Y}_i&=\sbrackets{\gvec{y}_1^{\brackets{i}},\ldots,\gvec{y}_{T_i}^{\brackets{i}}},  
\hspace{0.5em} 
\gvec{X}_i=\sbrackets{\gvec{x}_0^{\brackets{i}},\gvec{x}_1^{\brackets{i}},\ldots,\gvec{x}_{T_i}^{\brackets{i}}},
\end{align} 
and $T_i$ is the length of the $i$th training trajectory. The training procedure aims at minimizing the regularized $\ell_2$ loss. Letting $\NNParam$ be the trainable parameters of the \ac{rnn} and $\hat{\gvec{x}}_t^{\brackets{i}}\brackets{\NNParam}$ be the output of \acl{kn} with parameters $\NNParam$ at time $t$ applied to $\gvec{Y}_i$, the loss is computed for the $i$th trajectory as 
\begin{align} 
\label{eqn:SupervisedLoss}
l_i(\NNParam) = \frac{1}{T_i}\sum_{t=1}^{T_i}\norm{
\hat{\gvec{x}}_t^{\brackets{i}}\brackets{\NNParam}\!-\! \gvec{x}^{\brackets{i}}_t}^2
+\gamma\cdot\norm{\NNParam}^2,
\end{align}
where $\gamma>0$ is a regularization coefficient. The loss measure \eqref{eqn:SupervisedLoss} is used to optimize $\NNParam$ via \ac{sgd} optimization combined with the \ac{bptt} algorithm \cite{werbos1990backpropagation}. 

%
\section{Unsupervised KalmanNet}\label{sec:KalmanNet}
%
%
%
\subsection{Unsupervised Training Algorithm} \label{ssec:training}
\acl{kn}, described in Subsection~\ref{ssec:supervised}, as well as other previously proposed \ac{dnn}-based state estimators such as \cite{coskun2017long}, are designed to estimate $\gvec{x}_t$, and thus are trained so that the output approaches the ground-truth hidden state sequence. {\acl{kn}} admits an interpretable architecture owing to its hybrid \ac{mb}/\ac{dd} design, which preserves the flow of the \ac{kf}. We exploit this fact to propose a training algorithm for \acl{kn} that does not rely on ground-truth labels. 

{\bf Unsupervised loss: }
\acl{kn} uses its state estimates to predict the next observation via \eqref{eqn:obs} as an internal feature. While the accuracy of this prediction, e.g., the squared magnitude of the innovation process \eqref{eqn:innov}, depends on the accuracy in estimating $\gvec{x}_t$ and the observation noise, \eqref{eqn:innov} can be computed based solely on the observed sequence. Consequently, one can adapt \acl{kn} in an unsupervised manner by training it to minimize $\norm{\Delta \gvec{y}_t}^2$. This quantity can be used to compute the gradient with respect to the parameters of the \ac{rnn}, which outputs the learned \ac{kg}, by the derivative chain rule. Indeed,
\begin{align}
&\frac{\partial \|\Delta \gvec{y}_t\|^2 }{\partial\Kgain_{t-1}}
\stackrel{(a)}{=}\frac{\partial}{\partial\Kgain_{t-1} } \norm{\gvec{H}\cdot\gvec{F}\cdot\Kgain_{t-1}\cdot\ino{t-1} -\ino{t}^-}^2 \notag \\
&=2\cdot\gvec{H}^\top\cdot\gvec{F}^\top\cdot\brackets{\Kgain_{t-1}\cdot \ino{t-1} - \ino{t}^-}\cdot\ino{t-1}^\top,
\label{eqn:gradient1unsup}
\end{align}
where $\ino{t}^-\triangleq \gvec{y}_t -\gvec{H}\cdot\gvec{F}\cdot\hat{\gvec{x}}_{t-1\given{t-2}}$.
In \eqref{eqn:gradient1unsup}, $(a)$ holds since
\begin{align}
\hat{\gvec{y}}_{t\given{t-1}}&=\gvec{H}\cdot\gvec{F}\cdot\hat{\gvec{x}}_{t-1\given{t-1}} \notag \\
&=
\gvec{H}\cdot\gvec{F}\cdot\brackets{\hat{\gvec{x}}_{t-1\given{t-2}}+\Kgain_{t-1}\cdot\ino{t-1}}.
\end{align} 

The gradient in  \eqref{eqn:gradient1unsup} indicates that the $\ell_2$ norm of the innovation process can be used to learn the computation of the \ac{kg}, which involves the trainable parameters of \acl{kn} $\NNParam$. Similarly to \eqref{eqn:SupervisedLoss}, the resulting loss for the $i$th trajectory is 
\begin{align}
\label{eqn:UnsupervisedLoss}
\tilde{l}_i(\NNParam) = \frac{1}{T_i}\sum_{t=1}^{T_i}\norm{
\hat{\gvec{y}}_{t\given{t-1}}^{\brackets{i}}\brackets{\NNParam}\!-\! \gvec{y}^{\brackets{i}}_t}^2
+\gamma\cdot\norm{\NNParam}^2.
\end{align}
Unsupervised \acl{kn} is thus trained using solely observed trajectories based on the loss measure \eqref{eqn:UnsupervisedLoss} using \ac{sgd} variants combined with \ac{bptt} for gradient computation.

{\bf Offline versus online training: } 
The ability to train \acl{kn} without providing ground-truth state sequences gives rise to two possible training approaches: a purely unsupervised offline training scheme, and an online semi-supervised strategy. The offline approach follows conventional unsupervised learning using unlabeled data of the form $\tilde{\mathcal{D}}= \set{\brackets{\gvec{Y}_i}}_{i=1}^N$. This data set is used to optimize $\NNParam$ via mini-batch \ac{sgd}-based optimization, where for every {batch} indexed by $k$, we choose $M < N$  trajectories indexed by $i_1^k, \ldots, i_M^k$, computing the mini-batch loss as
%
$\mathcal{L}_k\brackets{\NNParam}=
\frac{1}{M}\sum_{j=1}^M\tilde{l}_{i_j^k}\brackets{\NNParam}$.

Online training builds upon the ability to learn without labels to adapt a pre-trained \acl{kn} to dynamics that differ from those used during training. Pretraining can be done using labeled data obtained by mathematical modelling and/or past measurements without altering the architecture of \acl{kn}. Then,  the deployed model is further adapted in an unsupervised manner using observations acquired during operation to form training trajectories from realizations of the data. Such a training procedure provides \acl{kn} with the ability to be adaptive to changes in the distribution of the data. {Specifically}, once every $\tilde{T}$ time steps, we compute the loss \eqref{eqn:UnsupervisedLoss} over the last $\tilde{T}$ observations online and optimize the \ac{rnn} parameters $\NNParam$ accordingly.
\subsection{Discussion}\label{subsec:discussion}
The ability to train \acl{kn} in an unsupervised manner without relying on ground-truth   sequences, follows directly from the  hybrid \ac{mb}/\ac{dd} architecture of \acl{kn}, where one can identify the observations innovation process as an internal feature and use it to compute the loss. The training procedure does not affect the \acl{kn} architecture, and one can use the same supervised model designed in \cite{KalmanNetTSPa}.
In the current work, we focus on partially-known \ac{ss} models  where $\gvec{F}$ and $\gvec{H}$ are available from, e.g., a physical model. While  supervised \acl{kn} was shown in \cite{KalmanNetTSPa} to operate reliably when using inaccurate approximations of $\gvec{F}$ and $\gvec{H}$, we leave such a study in unsupervised setups for future work.

The proposed online semi-supervised technique allows one to adapt a pre-trained \acl{kn} state estimator after deployment, coping with setups in which the original training is based on data that does not fully capture the true underlying \ac{ss} model. This gain, numerically demonstrated in Section~\ref{sec:NumEval}, bears some similarity to online training mechanisms proposed for hybrid \ac{mb}/\ac{dd} communication receivers in  \cite{shlezinger2019viterbinet, shlezinger2019deepSIC,teng2020syndrome}. Despite the similarity, the proposed technique,   obtained from the interpretable operation of \acl{kn}, is fundamentally different from that proposed in \cite{shlezinger2019viterbinet, shlezinger2019deepSIC,teng2020syndrome}, where  structures in communication data were exploited to generate confident labels from decisions.  Nonetheless, both the current work and \cite{shlezinger2019viterbinet, shlezinger2019deepSIC,teng2020syndrome} demonstrate the potential of \ac{mb} deep learning  in enabling application-oriented, efficient training algorithms. 

%
%
\section{Numerical Evaluations}\label{sec:NumEval}
{In this section} we numerically\footnote{{The source code used in our numerical study along with the complete set of hyper-parameters used in each numerical evaluation can be found online at \url{https://github.com/KalmanNet/Unsupervised_ICASSP22}.}} evaluate unsupervised \acl{kn} on a linear \ac{ss} model and on the \acl{nl} \acl{la}  model, and compare it to the \ac{kf} and extended \ac{kf}.

\begin{figure}
\centering
\includegraphics[width=1\columnwidth]{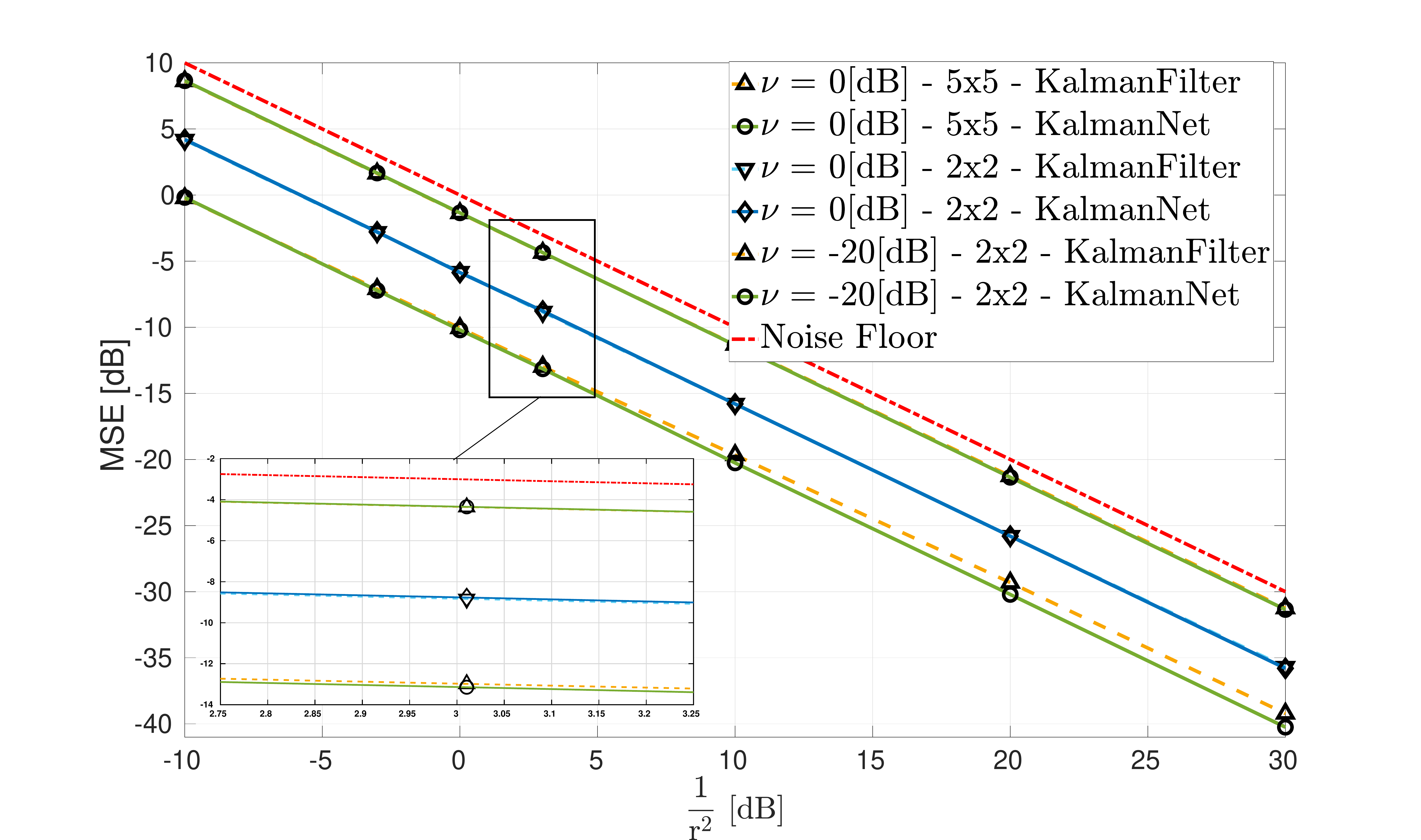}
\vspace{-0.6cm}
\caption{\ac{mse} vs $\frac{1}{\gscal{r}^2}$ for a $2\times2$ and a $5\times5$ linear systems, $T=80$.  An \ac{mse} offset is added to prevent overlapping.}
\label{fig:simLinear}
\end{figure}

In the linear setup, $\gvec{F}$ and $\gvec{H}$ take a canonical form, and $\gvec{Q}$ and $\gvec{R}$ are the diagonal matrices $\gscal{q}^2\cdot\gvec{I}_m$ and $\gscal{r}^2\cdot\gvec{I}_n$, respectively, while defining $\nu\triangleq\frac{\gscal{q}^2}{\gscal{r}^2}$. In Fig.~\ref{fig:simLinear} we compare the performance of unsupervised \acl{kn} to the \ac{mb} \ac{kf}, which achieves the \ac{mmse} here, for $2\times 2$ and $5\times 5$ \ac{ss} models and trajectory length  $T=80$. We observe in Fig.~\ref{fig:simLinear} that the \emph{offline} trained unsupervised \acl{kn} learns to achieve the \ac{mmse} lower bound. 
Next, the previously trained model is evaluated on a longer trajectory length of $T=10,000$. The results reported in Table~\ref{tbl:gen_T} show that \acl{kn} does not overfit to the trajectory length and that the unsupervised training of \acl{kn} is not tailored to the trajectories presented during training, tuning the filter with dependency only on the \ac{ss} model.
%
%
%
%
\begin{table}
\begin{center}
{\scriptsize
\begin{tabular}{|c|c|c|c|c|c|c| }
\hline
$\frac{1}{\gscal{r}^2}\dB$ & $0$ & $3$ & $10$ & $20$ & $30$\\  
\hline
\ac{kf} \ac{mse} $\dB$ & $-2.31$ & $-5.31$ & $-12.3$ & $-22.3$ & $-32.3$ \\
\hline
\acl{kn} \ac{mse} $\dB$ & $-2.27$ & $-5.26$ & $-12.3$ & $-22.3$ & $-32.3$ \\
\hline
\end{tabular}
}
\end{center}
\vspace{-0.2cm}
\caption{Generalizing for long trajectories, $2\times2$, $\nu=0\dB$. }
\label{tbl:gen_T}
\vspace{-0.2cm}
\end{table} 

\begin{figure}
\centering
\includegraphics[width=1\columnwidth]{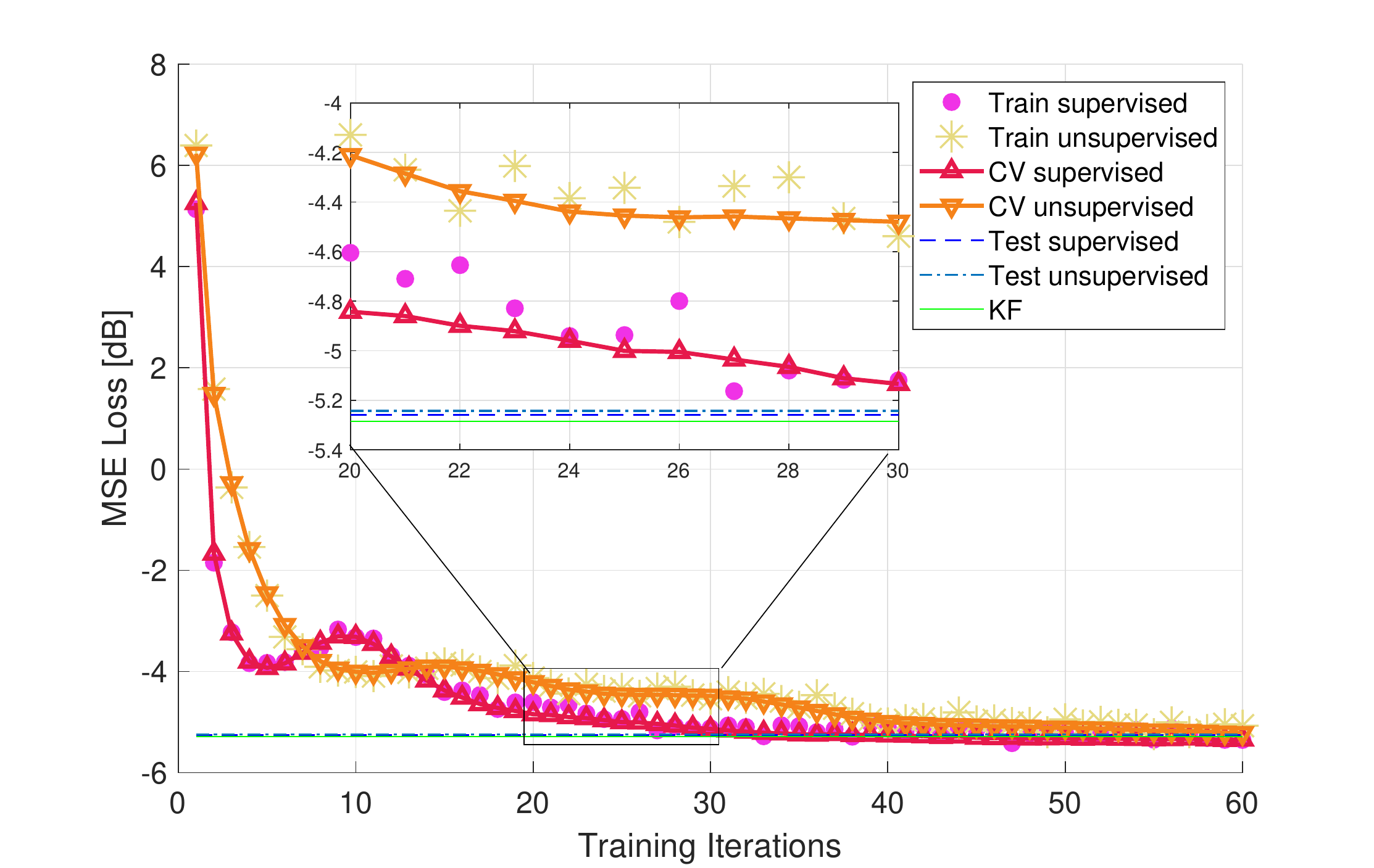}
\vspace{-0.4cm}
\caption{\ac{mse} versus epoch of supervised and unsupervised \acl{kn} for a $2\times2$ linear \ac{ss} model, $\nu=0\dB$.}
\label{fig:convergence_gt}
\vspace{-0.4cm}
\end{figure}

The results so far indicate that for the considered \ac{ss} model, unsupervised training  does not degrade the performance of \acl{kn} observed for supervised training in \cite{KalmanNetTSPa}. To understand the benefits of supervision, we depict in Fig.~\ref{fig:convergence_gt} the \ac{mse} convergence of unsupervised \acl{kn} compared with its supervised counterpart. We observe in Fig.~\ref{fig:convergence_gt} that the lack of labeled data in unsupervised \acl{kn} and the fact that it is not explicitly encouraged to minimize the state estimation \ac{mse}, results in slower convergence to the \ac{mmse}  compared to supervised \acl{kn}. 
%
%

\begin{figure}
\centering
\includegraphics[width=1\columnwidth]{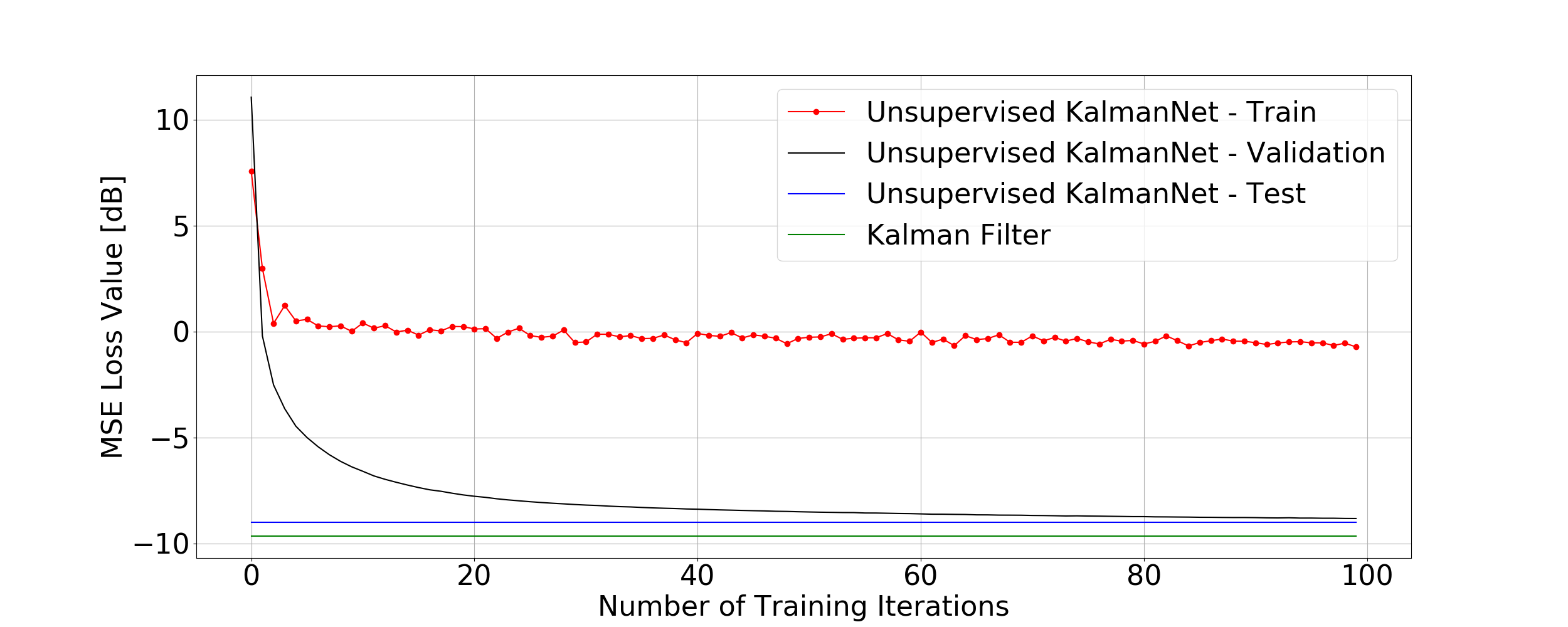}
\vspace{-0.4cm}
\caption{Learning curve of unsupervised \acl{kn}, \acl{la}. $\gscal{r}^2=0\dB$, $\gscal{q}^2=0\dB$, $T=100$.}
\vspace{-0.4cm}
\label{fig:non-linear}
\end{figure}

Next, we train unsupervised \acl{kn} for the  \acl{nl} \ac{ss} model of the chaotic \acl{la} (see \cite{KalmanNetTSPa} for details).
In Fig.~\ref{fig:non-linear} we can see that we were able to train \acl{kn} for this challenging setup. Although the training test is bounded by the observation noise $\gscal{r}^2=0\dB$, the \ac{mse} achieved by unsupervised \acl{kn} is within a minor gap of $0.5\dB$ from the \ac{mse} achieved by the extended \ac{kf} which has full knowledge of the \ac{ss} model. Furthermore, the \ac{dnn}-aided \acl{kn} is observed to require $0.15$ seconds to infer for each trajectory, which is quicker compared to the extended \ac{kf}, that involves matrix inversions and requires $0.2$ seconds per trajectory. This indicates that \acl{kn} may be preferable even when one can  estimate the noise statistics. 
%

\begin{figure}
\centering
\includegraphics[width=1\columnwidth]{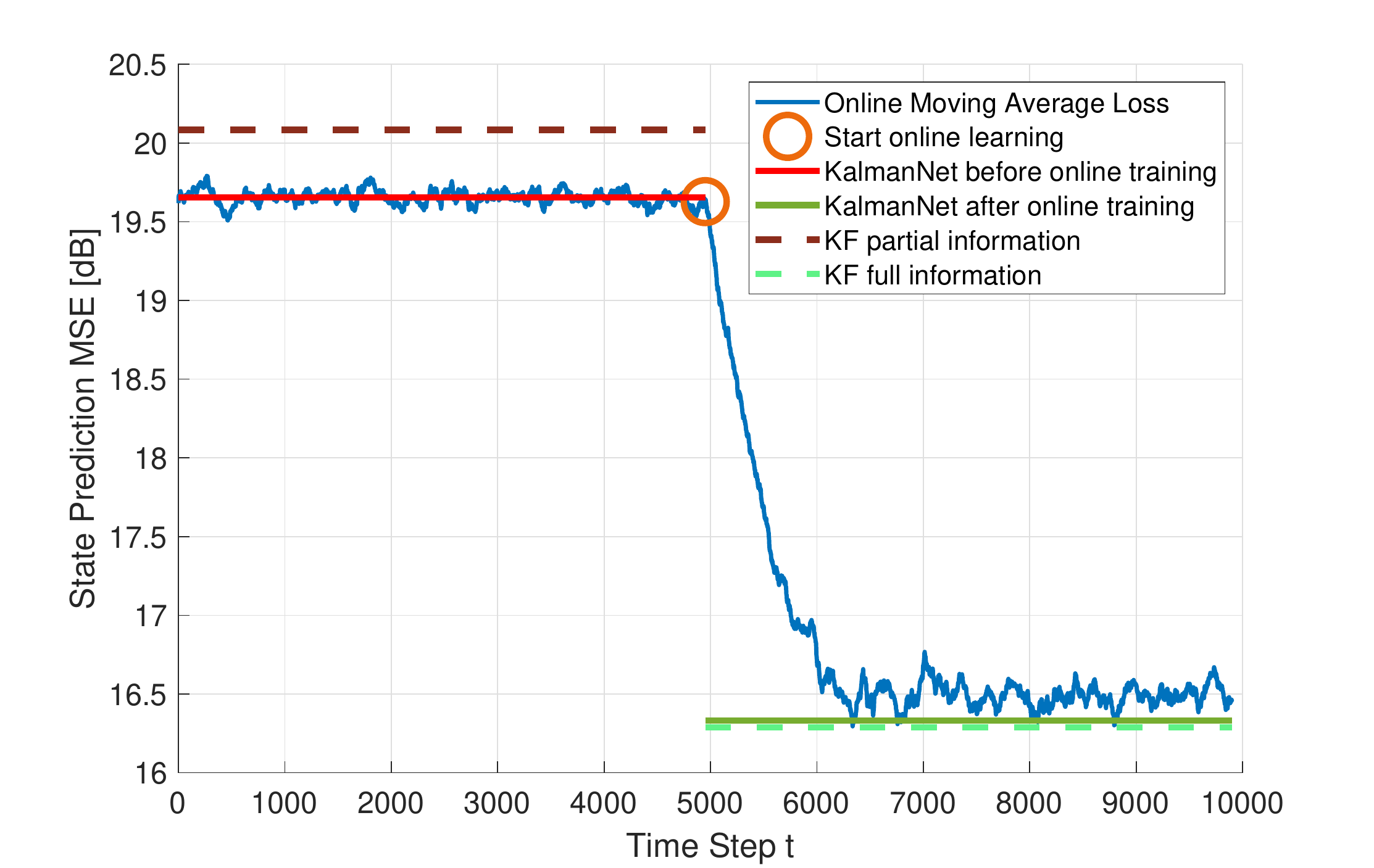}
\vspace{-0.4cm}
\caption{A pre-trained \acl{kn} trained in an {online} unsupervised manner. $2\times2$ linear \ac{ss} model, $\gscal{q}^2=10\dB$}
\vspace{-0.4cm}
\label{fig:online}
\end{figure}

%
Finally, we evaluate the online training mechanism when the testing distribution differs from the \ac{ss} model from which the training data is generated. We again consider a linear $2\times2 $ \ac{ss} model where the true (testing) observation distribution is generated with $\gscal{r}^2=25\dB$, while the model is pre-trained on data from an \ac{ss} model with $\gscal{r}^2=10\dB$. For online adaptation, we train every incoming $\tilde{T}=10$ samples. In Fig.~\ref{fig:online} we can see that \acl{kn} smoothly adapts to the test distribution while training on the observed trajectory over multiple time steps. This shows that the proposed training algorithm enables \acl{kn} to track variations in the \ac{ss} model.
%
%
\section{Conclusions}\label{sec:Conclusions}
In this work we proposed an unsupervised training scheme that enables  \acl{kn}  to learn its mapping without requiring ground-truth sequences. The training scheme exploits the interpretable nature of \acl{kn} to formulate an unsupervised loss  based on an internal feature that predicts the next observation. Our numerical evaluations demonstrate that the proposed unsupervised training allows \acl{kn} to approach the \ac{mmse}, without access to the noise statistics.

%
\bibliographystyle{IEEEtran}
\bibliography{IEEEabrv,KalmanNet}
\end{document}